\documentclass[12pt]{article}
\usepackage{psfig}
\def\Div{{\rm Div}}
\def\Rot{{\rm Rot}}
\def\pp{{\overline\partial}}

\def\stTD#1#2{\hbox to 0em{\mathsurround=0em %
$\stackrel{#1}{#2}$\hss} \phantom{#2}}
\def\stscript#1#2{\hbox to 0em{\mathsurround=0em %
${\scriptstyle\stackrel{#1}{#2}}$\hss} \phantom{#2}}
\def\stscriptscript#1#2{\hbox to 0em{\mathsurround=0em %
${\scriptscriptstyle\stackrel{#1}{#2}}$\hss} \phantom{#2}}
\def\st#1#2{\mathchoice{\stTD{#1}{#2}}{\stTD{#1}{#2}}%
{\stscript{#1}{#2}}{\stscriptscript{#1}{#2}}}

\def\bt{\begin{tabular}}
\def\bwt{\begin{tabular}{lcl}}
\def\et{\end{tabular}}
\def\bd{\bar{d}}

\def\rn#1{\st{#1}{r}}
\def\dn#1{\stackrel{#1}{d}}
\def\bn#1{\stackrel{#1}{b}}

\def\uz{\underline{z}}

\def\bfgr#1{\mbox{{\boldmath $#1$}}}
\def\jm{\jmath}
\def\bjm{\bar{\jmath}}
\def\bfjm{{\bfgr \jmath}}
\def\bfbjm{\bar{\bfgr \jmath}}

\def\d{{\rm d}}
\def\x{{\bf x}}
\def\p{\partial}

\def\be{\begin{equation}}
\def\ee{\end{equation}}
\def\ba#1{\begin{array}{#1}}
\def\ea{\end{array}}
\def\bea{\begin{eqnarray}}
\def\eea{\end{eqnarray}}

\def\dfrac#1#2{{\displaystyle\frac{#1}{#2}}}
\def\dint#1#2{{\displaystyle\int\limits^{#1}_{#2}}}

\def\E{{\bf E}}
\def\B{{\bf B}}
\def\D{{\bf D}}
\def\H{{\bf H}}

\def\Om#1{\st{#1}{\Omega}}

\def\bfV#1{\st{#1}{\bf V}}
\def\bV2#1{\st{#1}{\bf V}^2}

\def\bfa#1{\st{#1}{\bf a}}

\def\m#1{\st{#1}{m}}

\def\const{{\rm const}}
\def\sts#1{\st{#1}{\sigma}}

\def\m0{0}

\def\cE{{\cal E}}

\def\cH{{\cal H}}

\def\cP{{\cal P}}
\def\cbP{{\bfgr{\cal  P}}}

\newcommand{\bref}{\begin{thebibliography}{99}}
\newcommand{\eref}{\end{thebibliography}}

\begin{document}
%\setlength{\textheight}{7.7truein}  %for 2nd page onwards

%\runninghead{A.A. Chernitskii} {Bidyon or an electromagnetic
%model for charged particle with spin}

\thispagestyle{empty}

\title{\bf Bidyon as an electromagnetic\\ model for charged particle with spin}%
\author{{\bf Alexander A. Chernitskii}\\[0.5ex]
\small St.-Petersburg Electrotechnical University,
Prof. Popov str. 5\\
\small St.-Petersburg, 197376, Russia;\\[0.4ex]
\small A.~Friedmann Laboratory for Theoretical Physics,
St.-Petersburg\\[0.4ex]
{\small \tt aa@cher.etu.spb.ru}}%
\date{\small Received October 16, 2002; Revised February, 2003}
\maketitle
%\thanks{}%
%\subjclass{}%
%\keywords{}%

%\date{}%
%\dedicatory{}%
%\commby{}%
% ----------------------------------------------------------------
\begin{abstract}
A general model of nonlinear electrodynamics with dyon
singularities is considered. We consider the field configuration
having two dyon singularities with identical electric and opposite
magnetic charges and we name it bidyon. We investigate the sum
of two dyon solutions as an initial approximation to the bidyon
solution. We consider the case when the velocities of the dyons
have equal modules and opposite directions on a common line. It is
shown that the associated field configuration has a constant full
angular momentum which is independent of distance between the
dyons and their speed. This property permits a consideration of this
bidyon configuration as an electromagnetic model for charged
particle with spin. We discuss the possible electrodynamic world
with oscillating bidyons as particles.
\end{abstract}
% ----------------------------------------------------------------
\newpage
A hypothetical particle having both electric and magnetic charges
is said to be dyon \cite{Schwinger}. An electromagnetic field
configuration with $N$ point dyons satisfies the following two
differential conditions:
\be
\label{Eq:Maxwella} \left\{ \ba{rcl} \Div\D &=& 4\pi\,\jm^0
\\[1ex]
\Div\B &=& 4\pi\,\bjm^0 \ea \right. \quad, \ee where\quad $\Div\D
{\,}\equiv{\,} \pp_i\,D^i$\quad,\qquad $\pp_\mu {\,}\equiv{\,}
\dfrac{1}{\sqrt{|g|}}\,\dfrac{\p}{\p x^\mu}\,\sqrt{|g|}$
\quad,\qquad $g {}\equiv{} \det(g_{\mu\nu})$\quad,
\\[0.8ex]
the Latin indices take the values $1,2,3$, the Greek ones take
the value $0,1,2,3$, $g_{\mu\nu}$\quad is a metric of
space-time coordinate system,
\be
\label{Def:current0}
\jm^0 {}\equiv{} \dfrac{1}{\sqrt{|g|}}\,{\displaystyle
\sum\limits_{n {}={} 1}^N}\,\dn{n}\, \delta(\x {}-{} \bfa{n})
\quad,\qquad \bjm^0 {}\equiv{}
\dfrac{1}{\sqrt{|g|}}\,{\displaystyle \sum\limits_{n {}={}
1}^N}\,\bn{n}\,\delta(\x {}-{} \bfa{n}) \quad,
\ee
$\dn{n}$ is an electric charge and $\bn{n}$ is a magnetic one for $n$-th
singular point,
 $\bfa{n} {}={} \bfa{n}(x^0)$ is a trajectory of it.

%\noindent
Here we use the definition for three-dimensional
$\delta$-function which is suitable for discontinuous functions
$f(\x)$:
\begin{eqnarray}
\int\limits_{\Om{n}} f(\x)\,\delta(\x {}-{} \bfa{n})\,(\d x)^3
{}\equiv{} \lim_{{\sts{n}} \to 0}\,\Biggl[\frac{1}{|\sts{n}|}\,
\int\limits_{{\sts{n}}} f(\x)\;\d{{\sts{n}}} \Biggr] \quad,\qquad
|\sts{n}|  {}\equiv{}  \int\limits_{{\sts{n}}}\d{{\sts{n}}} \quad,
\label{Def:delta}
\end{eqnarray}
where $\Om{n}$ is a region of three-dimensional space including
the point $\x  {}={}  \bfa{n}$,
 $\sts{n}$ is a closed surface enclosing this point,
$\d{{\sts {n}}} $ is an area element of the surface $\sts{n}$,
$|\sts{n}|$ is an area of the whole surface $\sts{n}$.

Eqs. (\ref{Eq:Maxwella}) are the part of Maxwell system of
equations in any space-time coordinate system with a metric
$g_{\mu\nu}$ (see also \cite{I1998HPA}).

To have a natural interaction between the dyons
(see \cite{I1998HPA,Idyonint}) we must take the
fields $\D$ and $\B$ satisfying some nonlinear Maxwell equations
that may be written in the following general form:
\be
\label{Eq:Maxwellb} \left\{ \ba{rcl} \pp_0 \,\D  {}-{}  \Rot\H  &=&
-4\pi\,\bfjm
\\[1ex]
\pp_0 \,\B  {}+{}  \Rot\E  &=& -4\pi\,\bfbjm \ea \right. \quad,
\ee where\quad $(\Rot \E)^i {}\equiv{} -\varepsilon^{0ijk}\,\p
E_k/\p x^j\;$, \quad $\varepsilon^{0123} {}={} -|g|^{-1/2}\;$,
\quad $\varepsilon_{0123} {}={} |g|^{1/2}\;$, \bea \label{Def:EH} &&
E_i {}={} \dfrac{\p \cH}{\p D^i} \quad,\qquad H_i {}={} \dfrac{\p
\cH}{\p B^i} \quad,\qquad \cH {}={} \cH (\D,\B) \quad,
\\
&& \bfjm {}\equiv{} \dfrac{1}{\sqrt{|g|}}\,{\displaystyle
\sum\limits_{n {}={} 1}^N}\,\dn{n}\,\bfV{n}\, \delta(\x {}-{}
\bfa{n}) \,,\; \bfbjm {}\equiv{}
\dfrac{1}{\sqrt{|g|}}\,{\displaystyle \sum\limits_{n {}={}
1}^N}\,\bn{n}\,\bfV{n}\, \delta(\x {}-{} \bfa{n}) \,,\;
 \bfV{n} {}\equiv{} \dfrac{\d \bfa{n}}{\d x^0}\,.\quad
\label{Def:currentV}
\eea

%\vspace{1.5ex}
According to Eqs. (\ref{Def:EH}) we have some dependencies
 $\E {}={} \E (\D,\B)$ and $\H {}={} \H (\D,\B)$
(see also \cite{BialynickiBirula,Idyonint}).
If $\D$, $\B$ appears only quadratically in $\cH$
then we have a linear electrodynamics but in general case
the function $\cH (\D,\B)$ defines a nonlinear electrodynamic model.
In this approach the
vector fields $\D$, $\B$ play the role of unknown functions for
system of equation (\ref{Eq:Maxwellb}) with additional
differential conditions (\ref{Eq:Maxwella}). This representation
is best suitable for an investigation of the interaction between the
dyons. From the fields $\E$, $\B$ satisfying equations
(\ref{Eq:Maxwella}), (\ref{Eq:Maxwellb}) we can obtain an
appropriate electromagnetic potential. In the case of the dyon singularity
of electromagnetic field a space part of the four-potential has
a line singularity \cite{Idyonint}.

The singular currents (\ref{Def:current0}), (\ref{Def:currentV}) must satisfy
to the following condition \cite{Idyonint}:
\begin{eqnarray}
\label{Cond:Fjfbj} F_{\mu\nu}\,\jm^\nu {}-{}
\dfrac{1}{2}\,\varepsilon_{\mu\nu\sigma\rho}\,f^{\sigma\rho}\,\bjm^\nu
&=& 0 \quad,
\end{eqnarray}
where $F_{i0}=E_i$, $F_{ij}=\varepsilon_{0ijk}\, B^k$, $ f^{0i}=D^i$, $f^{ij} = \varepsilon^{0ijk}\, H_k$.

Using Eqs. (\ref{Eq:Maxwella}), (\ref{Eq:Maxwellb}),
(\ref{Def:EH}), (\ref{Cond:Fjfbj}) we can check directly the
following differential conservation laws for
%densities of field
energy-momentum tensor (in Cartesian coordinate systems):
\bea
\label{Def:ConscH} \frac{\p \cH}{\p x^0} &=& -\frac{\p}{\p x^j}
\left(\varepsilon^{jpq}\,E_p\,H_q \right)
\quad,
\\
\frac{\p \cP_i}{\p x^0}  &=& -\frac{\p}{\p x^j} \left[ \delta_i^j
\left(\D\cdot\E {}+{} \B\cdot\H {}-{} \cH \right) {}-{}
\left(D^j\,E_i {}+{} B^j\,H_i\right) \right] \label{Def:ConscP}
\quad,
\eea
where\quad $\cP_i {\,}\equiv{\,} \varepsilon_{ipq}\,D^p\,B^q$\quad
or\quad $\cbP {\,}\equiv{\,} \D {}\times{} \B$\quad
($\varepsilon_{123} {}={} \varepsilon^{123} {}={} 1$).

\vspace{1ex} From (\ref{Def:ConscH}) and (\ref{Def:ConscP}) we
easily obtain that the full energy-momentum\footnote{Note, here
we take the function $\cH$ such that $\cH {}={} 0$ for $\D {}={}
\B {}={} 0$. This is distinction from the designation which is
used in the article \cite{Idyonint} for Born-Infeld
electrodynamics.}
\begin{eqnarray}
\label{Def:EnergyMomentum}
\begin{array}{rclcrcl}
{\cE}&=&\dfrac{1}{4\pi} \dint{}{} \cH\;(\d x)^3 &\quad,\qquad&
{\bf P} &=& \dfrac{1}{4\pi} \dint{}{} \cbP\;(\d x)^3
\quad,
\end{array}
\end{eqnarray}
and the vector of full angular momentum
\begin{eqnarray}
\label{Def:VectAngMom} {\bf M} &=& \frac{1}{4\pi}\dint{}{} \left(
\x {}\times{} \cbP \right) (\d x)^3
\end{eqnarray}
are conserved on time, i.e. $\d \cE/\d x^0 {}={} \d {\bf P}/\d x^0
{}={} \d {\bf M}/\d x^0 {}={} 0$.

Let us consider a solution of system (\ref{Eq:Maxwellb}),
(\ref{Eq:Maxwella}) having two dyon singularities with identical
electric and opposite magnetic charges: $\dn{1} {}={} \dn{2}$,
$\bn{1} {}={} -\bn{2}$. We name this solution bidyon. Let us
consider the case when the velocities of the singularities have
equal absolute values and opposite directions on a common line. At
first we use a cylindrical coordinate system $\{z,\rho,\varphi \}$
such that the dyon singularities
 are on the axis $z$.
This configuration is shown in Fig.~1, where $d {}={} \pm\bar{d}$,
$b {}={} \pm\bar{b}$ and $\bar{d},\,\bar{b}$ are some positive
constants.

\begin{figure}[htbp]
\label{Fig:twodyons}
\begin{picture}(300,115)
\put(170,15){
\begin{picture}(100,100)
\put(0,0){\vector(0,1){92}} \put(-8,92){$z$}
\put(-2,45){\line(1,0){4}} \put(-10,42){$0$}
\put(0,15){\circle*{3}} \put(-18,13){$-a$}
\put(10,15){\vector(0,1){20}} \put(15,25){$\bfV{1} {}={}
{-\bfV{}}$} \put(15,5){$\dn{1} {}={} d,\;\,\bn{1} {}={} {-b}$}
\put(0,75){\circle*{3}} \put(-18,73){$\phantom{-}a$}
\put(10,75){\vector(0,-1){20}} \put(15,55){$\bfV{2} {}={} \bfV{}$}
\put(15,75){$\dn{2} {}={} d,\;\,\bn{2} {}={} {b}$} \put(-60,45){
\begin{picture}(50,50)
\put(0,0){\vector(-1,0){20}} \put(-24,-6){${\bf e}_\rho$}
\put(0,0){\vector(0,1){20}} \put(0,22){${\bf e}_z$}
\put(0,0){\vector(1,-1){10}} \put(10,-6){${\bf e}_\varphi$}
\end{picture}
}
\end{picture}
}
\end{picture}
\caption{Disposition of the two dyons in the cylindrical
coordinate system.}
\end{figure}
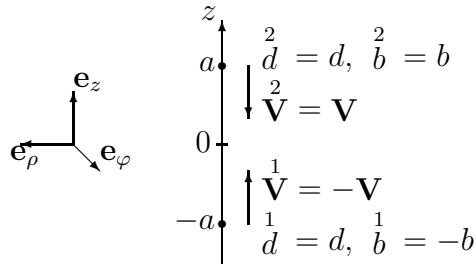

We can search the solution by some iterative procedure and we can
take a sum of two moving dyon solutions as initial
approximation. That is we consider the following initial approximation
to the bidyon solution:
\be
\label{Sol:InitBidyon} \D^{(0)} {}={} \st{1}{\D} {}+{} \st{2}{\D}
\quad,\qquad \B^{(0)} {}={} \st{1}{\B} {}+{} \st{2}{\B} \quad.
\ee
%where $(\st{1}{\D},\st{1}{\B})$, $(\st{2}{\D},\st{2}{\B})$ are two
%dyon solutions to system (\ref{Eq:Maxwellb}), (\ref{Eq:Maxwella}).
%For Born-Infeld model there is

\newpage
Here we consider the dyon solutions with constant velocity.
For $z$- and $\rho$-components of these solutions (see
\cite{Idyonint}) we have the following expressions:
\begin{eqnarray}
\label{BidyonFieldInitDB} \ba{l}
\begin{array}{lcl}
\left\{
\begin{array}{rcrcl}
\dfrac{\st{1}{D}_z}{d} &=& \dfrac{\st{1}{B}_z}{-b} &=&
\dfrac{1}{\sqrt{1 {}-{} V^2}}\,\dfrac{z {}+{} a}{\rn{1}^3}
\\[2ex]
\dfrac{\st{1}{D}_\rho}{d} &=& \dfrac{\st{1}{B}_\rho}{-b} &=&
\dfrac{1}{\sqrt{1 {}-{} V^2}}\, \dfrac{\rho}{\rn{1}^3}
\end{array}
\right. \quad,
\\[2.5ex]
\left\{
\begin{array}{rcrcl}
\dfrac{\st{2}{D}_z}{d} &=& \dfrac{\st{2}{B}_z}{b} &=&
\dfrac{1}{\sqrt{1 {}-{} V^2}}\,\dfrac{z {}-{} a}{\rn{2}^3}
\\[2ex]
\dfrac{\st{2}{D}_\rho}{d} &=& \dfrac{\st{2}{B}_\rho}{b} &=&
\dfrac{1}{\sqrt{1 {}-{} V^2}}\, \dfrac{\rho}{\rn{2}^3}
\end{array}
\right. \quad,
\end{array}
\ea
\end{eqnarray}
\bt{ll} where&\quad $V {}\equiv{} \dfrac{\d a}{\d x^0}$ ,\quad
$\rn{1} {}={} \sqrt{\left(z^\prime {}+{} a^\prime\right)^2 {}+{}
\rho^2 }$ ,\quad $\rn{2} {}={} \sqrt{\left(z^\prime {}-{}
a^\prime\right)^2{}+{} \rho^2 }$ ,\\[2ex] &\quad $z^\prime
{}\equiv{} \dfrac{z}{\sqrt{1 {}-{} V^2}} \quad,\qquad a^\prime
{}\equiv{} \dfrac{a}{\sqrt{1 {}-{} V^2}}$ \quad. \et\\[1ex]
Forms of $\varphi$-components for the vector fields $\st{1}{\D}$,
$\st{1}{\B}$, $\st{2}{\D}$, $\st{2}{\B}$ depend on forms of
the functions $\E {}={} \E (\D,\B)$, $\H {}={} \H (\D,\B)$ but
forms (\ref{BidyonFieldInitDB}) for $\rho$- and $z$-components are
independent of %the functions $\E {}={} \E (\D,\B)$, $\H {}={} \H (\D,\B)$
the specific model's nonlinearity. The
$\varphi$-components equals zero when $V {}={} 0$. The lines of
the vector fields $\D^{(0)}$ and $\B^{(0)}$ in $z\rho$-plane for
$V {}={} 0$ are shown in Fig.~2.
\begin{figure}[!ht]
\label{Fig:Fields}
\begin{picture}(330,180)
\put(5,-455){
%\special{bidyon9.eps}
\psfig{file=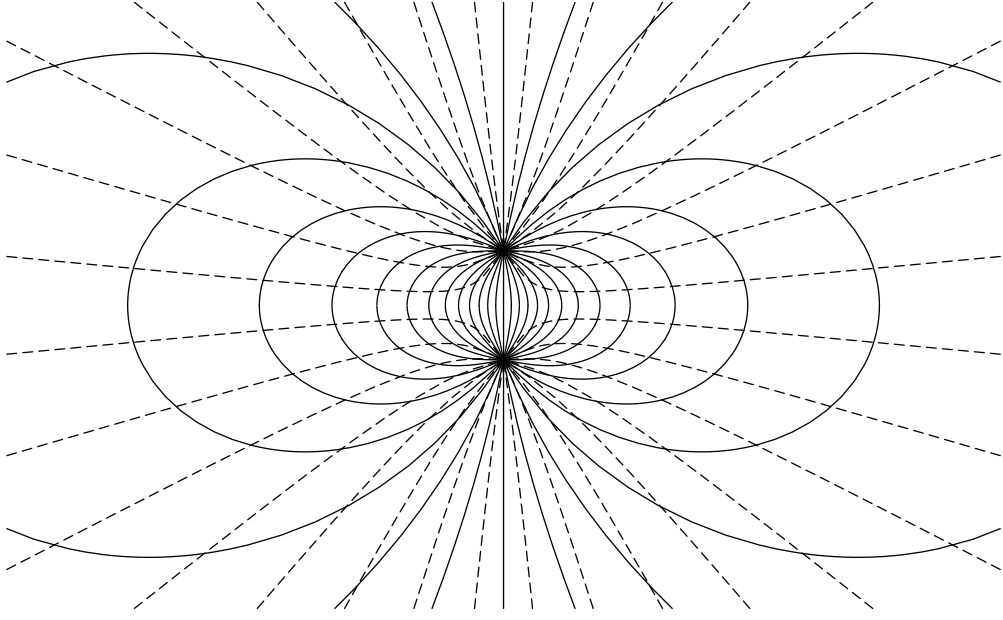,width=5.2truein} }
\end{picture}
\caption{Lines of the fields $\D^{(0)}$ (discontinuous lines) and
$\B^{(0)}$ (continuous lines).}
\end{figure}

Now let us calculate the vector of full
 angular momentum ${\bf M}$ (\ref{Def:VectAngMom})
for field configuration (\ref{Sol:InitBidyon}) with (\ref{BidyonFieldInitDB}).
Because of a symmetry property of the element of integration into
(\ref{Def:VectAngMom}), for our case we have $M_\rho {}={} M_\varphi {}={} 0$ and
\begin{eqnarray}
\label{Def:Mz} M_z &=& \frac{1}{4\pi} \int \cP^{(0)}_\varphi\,\rho
\;(\rho\, \d z \d\rho\d\varphi) \quad.
\end{eqnarray}

Using (\ref{BidyonFieldInitDB}) we can easily obtain the following
expression:
\begin{eqnarray}
\label{Expr:Pphi}
\cP^{(0)}_\varphi &=& \frac{4\,a\,b\,d\,\rho}%
{\rn{1}^3\,\rn{2}^3\,\left(1 {}-{} V^2\right)} \quad.
\end{eqnarray}

Substituting (\ref{Expr:Pphi}) into (\ref{Def:Mz}) and introducing
new variables of integration we obtain
\newpage
\begin{eqnarray}
\label{Calc:Mza} M_z &=& \frac{b\,d}{\pi \left(1 {}-{} V^2\right)}
\int \frac{a\,\rho^3}{\rn{1}^3\,\rn{2}^3}\, \d z \d\rho\d\varphi
\\
\label{Calc:Mzb} &=& \frac{b\,d}{\pi} \int
\frac{a^\prime\,\rho^3}{\rn{1}^3\,\rn{2}^3}\, \d z^\prime
\d\rho\d\varphi
\\
\label{Calc:Mzc} &=& \frac{b\,d}{\pi}
\int \frac{(\rho^{\prime\prime})^3}{(\rn{1}^{\prime\prime})^3\,%
(\rn{2}^{\prime\prime})^3}\, \d z^{\prime\prime}
\d\rho^{\prime\prime}\d\varphi \quad,
\end{eqnarray}
where\quad $z^{\prime\prime} {}\equiv{} z^\prime/a^\prime$
\quad,\qquad $\rho^{\prime\prime} {}\equiv{} \rho/a^\prime$
\quad,\\[1ex]
\phantom{where\quad} $\rn{1}^{\prime\prime} {}\equiv{} %
\sqrt{\left(z^{\prime\prime} {}+{} 1\right)^2%
 {}+{} \left(\rho^{\prime\prime}\right)^2 }$
\quad,\qquad
$\rn{2}^{\prime\prime} {}\equiv{} %
\sqrt{\left(z^{\prime\prime} {}-{} 1\right)^2%
 {}+{} \left(\rho^{\prime\prime}\right)^2 }$\quad.

\vspace{1.5ex}\noindent As we see, in first change
(\ref{Calc:Mza})$\to$(\ref{Calc:Mzb}) the dependence on speed is
canceled. In second change (\ref{Calc:Mzb})$\to$(\ref{Calc:Mzc})
the dependence on $a^\prime$ is canceled. Thus we obtain that the
full angular momentum for field configuration
(\ref{Sol:InitBidyon}), (\ref{BidyonFieldInitDB}) is independent
of dyon's speed ($V$) and distance between the dyons ($2\,a$)!

For calculation (\ref{Calc:Mzc}) we introduce the variables of
integration $\{\xi,\zeta,\varphi\}$ that appropriate to the
bispherical coordinate system with unit parameter characterizing positions of
focal points:
\begin{eqnarray}
\label{Trans:bispher} z^{\prime\prime} {}={}
\frac{\sinh{\xi}}{\cosh{\xi}{}-{} \cos{\zeta}} \quad,\qquad
\rho^{\prime\prime} {}={} \frac{\sin{\zeta}}{\cosh{\xi}{}-{}
\cos{\zeta}} \quad.
\end{eqnarray}
The bispherical element of value has the form
\begin{eqnarray}
\label{Expr:Value} (\rho^{\prime\prime}\, \d z^{\prime\prime}
\d\rho^{\prime\prime}\d\varphi) &=&
\dfrac{\sin{\zeta}\;\d\xi\,\d\zeta\,\d\varphi}%
{\left(\cosh{\xi}{}-{} \cos{\zeta}\right)^3} \quad.
\end{eqnarray}
We have also that
\begin{eqnarray}
\label{Expr:rr} \rn{1}^{\prime\prime} {}={}
\frac{\sqrt{2}\,\exp{\left(\xi/2\right)}}%
{\sqrt{\cosh{\xi}{}-{} \cos{\zeta}}} \quad,\qquad
\rn{2}^{\prime\prime} {}={} \frac{\sqrt{2}\,\exp{\left(-\xi/2\right)}}%
{\sqrt{\cosh{\xi}{}-{} \cos{\zeta}}} \quad.
\end{eqnarray}

Substituting (\ref{Trans:bispher}), (\ref{Expr:Value}),
(\ref{Expr:rr}) into (\ref{Calc:Mzc}) and introducing the variable
$\uz {}={} \cos{\zeta}$, we obtain
\begin{eqnarray}
M_z &=& \frac{b\,d}{8\pi}\, \int\limits^\pi_{-\pi} \d\varphi
\int\limits^\infty_{-\infty} \d\xi \int\limits^1_{-1}
\frac{\left(1 {}-{} \uz^2\right)}{\left(\cosh{\xi} {}-{}
\uz\right)^2} \;\d\uz \quad.
\end{eqnarray}
Making firstly the integration over $\uz$ and $\xi$ in the finite
limits $[-\bar{\xi},\,\bar{\xi}]$ we obtain
\begin{equation}
\int\limits^\infty_{-\infty} \d\xi \int\limits^1_{-1}
\frac{\left(1 {}-{} \uz^2\right)}{\left(\cosh{\xi} {}-{}
\uz\right)^2} \;\d \uz {\,}={\,} \lim_{\bar{\xi} {}\to{} \infty}
\left[ 4\left(\ln \frac{\cosh{\bar{\xi}} {}+{} 1}{\cosh{\bar{\xi}}
{}-{} 1} \right) \sinh{\bar{\xi}}\, \right] {\,}={\,} 8 \quad.
\end{equation}
Thus we have
\begin{eqnarray}
\label{Expr:Angmom} M_z &=& 2\,b\,d \quad.
\end{eqnarray}

Of course, the full angular momentum for an appropriate exact
solution is conserved. This implies that we may have internal
movements of the singularities, which don't change the full
angular momentum. Thus here we have verified that our choice of
two moving dyons (\ref{Sol:InitBidyon}), (\ref{BidyonFieldInitDB})
as the initial approximation is appropriate.

It is evident that the full momentum for field configuration
(\ref{Sol:InitBidyon}), (\ref{BidyonFieldInitDB}) is zero. To
verify satisfaction of the conservation law for full energy,
we must take a concrete function $\cH (\D,\B)$. That is, we must
investigate the concrete nonlinear electrodynamic model. In this
case the condition $\cE {}={} \const$ can be used for defining a trajectory
$a(x^0,\cE)$ of the dyons in the initial approximation.
This problem was investigated for Born-Infeld
electrodynamics \cite{Idyonint} and it was shown that the initial
field configuration may behave as nonlinear oscillator.
(A wave part of the dyon solutions connected with acceleration of the singular points
was not included to the initial approximation.)

The field configuration (\ref{Sol:InitBidyon}),
(\ref{BidyonFieldInitDB}) looks like charged particle with spin.
The charge of this particle is $2\,d$ and its spin is equal to
$|M_z|$. We may set $2\,\bd {}={} e$, where $e$ is the absolute
value of the electron charge, and $|M_z| {}={} \hbar {}/{} 2$. In
this case we have
\begin{eqnarray}
\bar{b}\,e {}={} \frac{\hbar}{2} \quad\Longrightarrow\quad \bar{b}
{}={} \frac{e}{2}\,\frac{\hbar}{e^2} {}={}
\frac{e}{2}\,\frac{1}{\bar{\alpha}} \quad\Longrightarrow\quad
\frac{\,\bd\,}{\,\bar{b}\,} {}={} \bar{\alpha} \quad,
\label{Def:baralpha}
\end{eqnarray}
\nopagebreak where $\bar{\alpha} {}={} e^2/\hbar {}\approx{}
1/137$ is the fine structure constant.

The field configuration  (\ref{BidyonFieldInitDB}) is considered
here as initial approximation to unknown exact bidyon solution.
This initial approximation does not include a wave part of the
bidyon solution. For periodical bidyon solution this wave part
must have the form of some standing wave localized near dyon
singularities. Because this problem has the boundary conditions
(which follow from (\ref{Cond:Fjfbj}),
see also \cite{Idyonint}) in two (moving) points (in which the dyon singularities are at
a current instant of time), it is possible that the bidyon solution has
some discrete set of allowable frequencies.

With the help of Lorentz transformation, from the rest oscillating
bidyon we can obtain an appropriate moving bidyon solution. It is
evident that the moving oscillating bidyon has both particle and
wave properties.

The field model under consideration allows existence for great
number of the dyon singularities with charges $\dn{n}$ and
$\bn{n}$. We may assume that there is some kind invariance of the
theory, such that
\begin{equation}
\dn{n} {}={} \pm \bar{d} \quad,\qquad \bn{n} {}={} \pm \bar{b}
\quad.
\end{equation}
For the suitable dimensional system we can take $\bar{d} {}={} 1$
or $\bar{b} {}={} 1$. Thus we have the relation $\bar{d}/\bar{b}$
as the single dimensionless constant of the theory. We can set
that this relation equals the fine structure constant
$\bar{\alpha}$ (\ref{Def:baralpha}).

We can fancy a world constructed from the great number of the
bidyon-type field configurations as particles. Particles with full
angular momentum
%multiple to
divisible by $\hbar {}/{} 2$ may be constructed from some number
of the bidyons. Because of $\bd {}/{} \bar{b} {}={} \bar{\alpha}
{}\ll{} 1$,
 we may build a perturbation
theory with the fine structure constant $\bar{\alpha}$ as small
parameter, for some aspects of the mathematical model of this
world. In this case we will have an analogy with the procedure of
perturbation theory in quantum electrodynamics.

As a result we can assume that there is some correlation between
the bidyon solution of a nonlinear electrodynamic model and
leptons.

% ----------------------------------------------------------------
%\bibliographystyle{amsplain}
%\bibliography{}
\end{document}